# Effective medium theory for anisotropic metamaterials

Xiujuan Zhang and Ying Wu[*]

*Division of Computer, Electrical and Mathematical Science and Engineering, King Abdullah University of Science and Technology (KAUST), Thuwal 23955-6900, Saudi Arabia*

Correspondence and requests for materials should be addressed to Y. W (ying.wu@kaust.edu.sa)

**Abstract:** We present an effective medium theory that can predict the effective permittivity and permeability of a geometrically anisotropic two-dimensional metamaterial composed with a rectangular array of elliptical cylinders. It is possible to obtain a *closed-form analytical* solution for the anisotropic effective medium parameters if the aspect ratio of the lattice and the eccentricity of elliptical cylinder satisfy certain conditions. The derived effective medium theory can recover the well-known Maxwell-Garnett results in the quasi-static regime. More importantly, it is valid beyond the long-wavelength limit, where the wavelength in the host medium is comparable to the size of the lattice so that previous anisotropic effective medium theories fail. The validity of the derived effective medium theory is verified by band structure calculations. A real sample of a recent theoretically proposed anisotropic medium with near-zero index to control flux is achieved from the derived effective medium theory and control of electromagnetic waves in the sample is clearly demonstrated. The accuracy dependence on the material and geometric parameters of the theory is further systematically discussed, and we find the derived effective medium theory applies for long range of parameters changing. Such applicability greatly broadens the applicable realm of the effective medium theory and opens many vistas in the design of structures with desired anisotropic material characteristics.

*Email: Ying.Wu@kaust.edu.sa

Metamaterials, which are artificially composed of subwavelength inclusions, can manipulate waves in surprising ways, thus exhibiting novel properties not occurring in nature, including negative refraction [1-3] and superlensing [4, 5]. One prominent class of metamaterials is anisotropic metamaterials [6], whose material parameters are not scalars but tensors with their principle components taking different values. Such property shapes elliptical or hyperbolic dispersion relations [7] in the anisotropic metamaterials, which promise distinctive properties including negative refraction [8, 9], super-resolution in the far-field through image magnification [10], and enhanced spontaneous emission [11]. Topological transitions occur [12-14] in anisotropic metamaterial when one principle component of its material parameter tensor changes its sign. Earlier this year, a theoretical method was proposed to arbitrarily control electromagnetic flux with a type of anisotropic media. In this medium, only one principle component is near zero and other components take positive values [15]. A real sample of the realization is yet to be reported.

It is known that the unconventional material parameters of a metamaterial are coming from the subwavelength nature of the structure and the local resonances of the building blocks: the subwavelength scale allows the heterogeneous materials to be homogenized as effective media, whereas local resonances lead to extraordinary effective medium parameters that are rarely or never found in nature. However, the existence of resonances will impose a big challenge to conventional effective medium theories (EMTs), such as the Maxwell-Garnett theory and the Bruggman theory [16], whose basic principle is to minimize the scattering at the quasi-static limit, while the local resonances usually happen in or even beyond the long-wavelength regime. The long-wavelength regime requires the wavelength in the host medium ($\lambda_0$) comparable to the size of the unit cell, and meanwhile, the wavelength in the scatterer ($\lambda_s$) can be very small [17].

In contrast, in quasi-static limit, both $\lambda_0$ and $\lambda_s$ should be much larger than the size of the unit cell. There have been many efforts to extend conventional EMTs towards higher frequency regimes. There was a coherent potential approximation method applied to both electromagnetic and elastic waves to enlarge the applicability range of the EMTs [18, 19]. Multiple-scattering formalism, which takes full account of the interactions among the scatterers, was also employed, and equivalent results were obtained [20, 21]. Based on Floquent representation, a rigorous approach was brought forward to homogenize metamaterials with periodic arrays of dielectric inclusions [22-24]. Later, a first-principles homogenization scheme was developed from dyadic Green's functions and polarizability coefficients, generally incorporating both dielectric and magnetic materials. An analytical solution was obtained for periodic systems with isotropic unit cells [25, 26]. Very recently, *Yang et al.* proposed a method based on reproducing the lowest orders of scattering amplitudes from a finite volume of metamaterials. Accurate predictions of the effective medium parameters almost over the whole Brillouin zone are obtained [27]. Notice that these schemes work well in isotropic media with both the scatterers and the lattice structures being isotropic. However, for anisotropic media, more challenges are added to the homogenization scheme as it involves more degrees of freedom than for isotropic media. Many conventional anisotropic EMTs are extensions of the Maxwell-Garnett theory [7, 28-30]; consequently they are limited to the application in quasi-static regime. Multiple-scattering-based schemes were introduced to study the effective medium properties of systems with anisotropic lattices and isotropic scatterers [31, 32], yielding scalar bulk modulus and tensorial mass density at finite frequencies in the long-wavelength regime. There also exist other schemes that can deal with anisotropic scatterers, such as the field-averaging [33, 34], boundary-integration [13, 35], and parameter-retrieving methods [36-39]. *Prior* knowledge of field distributions is required for the field-averaging and boundary-integration methods, while the information of transmission and reflection coefficients is required for the parameter-retrieving method and multiple solutions may be obtained. More importantly, none of these three methods can offer a *closed-form analytical* solution that is capable of predicting reliable effective medium parameters directly from the material and geometric information of the system.

In this work, a periodic rectangular array of elliptical cylinders embedded in air is considered. By studying its scattering properties, we find that the special properties of elliptical coordinates and Mathieu functions (solutions to Helmholtz equation in elliptical coordinates) promise a *closed-form analytical* solution for the anisotropic effective medium parameters if the aspect ratio and the eccentricity of the elliptical cylinder satisfy certain conditions. The derived EMT is verified by comparing the band structures from its predictions with the full-wave simulations. Excellent agreements are found at finite frequencies beyond the long-wavelength limit. This advantage suggests great opportunities to broaden the design of anisotropic metamaterials. Based on our EMT predictions, we show that a recently theoretically proposed anisotropic near-zero material, which can manipulate electromagnetic flux, can be achieved. Since the metamaterial is made of common dielectric materials with simple structures, the fabrication process will be feasible and the experimental realization of the material would be greatly benefited.

The system considered in our study is a two-dimensional (2D) metamaterial composed with a periodic rectangular array of elliptical cylinders with permittivity, $\varepsilon_s$, and magnetic permeability, $\mu_s$, embedded in a background material with permittivity, $\varepsilon_0$, and magnetic permeability, $\mu_0$. A unit cell of the metamaterial is illustrated in Fig. 1(a). $a_s$ and $b_s$ denote the elliptical cylinder's semi-major and semi-minor axes, respectively, and the filling ratio, i.e., the ratio of the area of the elliptical cylinder to the area of the unit cell, is $f$. With given $a_s$, $b_s$, and $f$, the periodicities, i.e. length $a$ and the width $b$ of the unit cell, are determined by $a^2 - b^2 = \pi\left(a_s^2 - b_s^2\right)$ and $abf = \pi a_s b_s$. In the dispersion microstructure [16], where the scatterers are always dispersed in the matrix, the EMT considers the scattering of a coated cylinder in an effective medium. As shown in Fig. 1(b), the inner elliptical cylinder represents the scatterer of the metamaterial and the coating layer is made of background medium. The semi-major and semi-minor axes of the outer coated elliptical cylinder are $a_0$ and $b_0$, respectively. To maintain the anisotropic property of the metamaterial and to produce a correct scattering property in the effective medium, both the aspect ratio and the cross-sectional area of the coated elliptical cylinder here should be the same as those of a rectangular unit cell, i.e., $a_0/b_0 = a/b$ and $\pi a_0 b_0 = ab$. The microstructure of the metamaterial is represented by the coated elliptical cylinder and the effective medium parameters $(\vec{\varepsilon}_{eff}, \vec{\mu}_{eff})$ are obtained when the total scattering of this cylinder vanishes. Here, because of the anisotropic nature of the metamaterial, we expect the effective medium parameters are tensors. Below, we present detailed steps to derive the EMT.

Consider the microstructure shown in Fig. 1(b). For a transverse-electric(TE)-polarized wave, in which the electric field is parallel to the elliptical cylinders ($\vec{E} = (0,0,E_z)$), the solution of the electric field in the effective medium can be expressed as [40]:

$$\begin{aligned} E_{ez} &= \sum_m \alpha_{em}(eff)ce_m(q_{eff};\eta)J_{em}(q_{eff};\xi) + \beta_{em}(eff)ce_m(q_{eff};\eta)H^{(1)}_{em}(q_{eff};\xi) \\ E_{oz} &= \sum_m \alpha_{om}(eff)se_m(q_{eff};\eta)J_{om}(q_{eff};\xi) + \beta_{om}(eff)se_m(q_{eff};\eta)H^{(1)}_{om}(q_{eff};\xi) \end{aligned}, \quad (1)$$

and similarly, the electric field in the background medium of the coating layer is:

$$E_{ez} = \sum_m \alpha_{em}(0) ce_m(q_0;\eta) J_{em}(q_0;\xi) + \beta_{em}(0) ce_m(q_0;\eta) H_{em}^{(1)}(q_0;\xi)$$

$$E_{oz} = \sum_m \alpha_{om}(0) se_m(q_0;\eta) J_{om}(q_0;\xi) + \beta_{om}(0) se_m(q_0;\eta) H_{om}^{(1)}(q_0;\xi)$$

(2)

Here, $\eta$ and $\xi$, with $0 \leq \eta < 2\pi$ and $0 \leq \xi < \infty$, represent the elliptical coordinates variables that can be transformed into Cartesian coordinates by using the chains $x = c\cos(\eta)\cosh(\xi)$ and $y = c\sin(\eta)\sinh(\xi)$, where $c = \sqrt{a_s^2 - b_s^2} = \sqrt{a_0^2 - b_0^2}$ is the focal length of the elliptical coordinate system. In Eqs. (1) and (2), $ce_m(q;\eta)$ and $se_m(q;\eta)$ denote the angular Mathieu functions of the first kind, while $J_m(q;\xi)$ and $H_m^{(1)}(q;\xi)$ are the radial Mathieu functions of the first kind and third kind, respectively. The subscript $m$ is an integer denoting the orders of the Mathieu functions. The angular and radial Mathieu functions form the solutions to the Helmholtz equation in elliptical coordinates, which split into two types of decoupled even modes (denoted by subscript $e$) and odd modes (denoted by subscript $o$) with respect to the $x$ axis for non-zero $m$. The variable $q_0$ ($q_{eff}$) is a dimensionless quantity and equals $\frac{1}{4}c^2 k_0^2$ ($\frac{1}{4}c^2 k_{eff}^2$), with $k_0 = \omega\sqrt{\varepsilon_0}\sqrt{\mu_0}$ ($k_{eff} = \sqrt{k_{eff,x}^2 + k_{eff,y}^2}$) being the wave vector in the background (effective) medium.

The expansion coefficients in Eqs. (1) and (2), i.e. $\alpha_{\gamma m}(\sigma)$ and $\beta_{\gamma m}(\sigma)$ with $\gamma = o, e$ and $\sigma = 0, eff$, are related through the boundary conditions, which are the continuities of both the electric field and tangential component of the magnetic field on the interface between the background medium and the effective medium. Such continuities can be expressed as $E_z(eff) = E_z(0)$ and $\partial_\xi E_z(eff)/\mu_{eff} = \partial_\xi E_z(0)/\mu_0$ at $\xi = \xi_0$, where $\xi_0$ is the outer boundary of the coated cylinder. Substituting Eqs. (1) and (2) into the boundary conditions, we obtain:

$$\begin{bmatrix} \alpha_{\gamma m}(eff) \\ \beta_{\gamma m}(eff) \end{bmatrix} = F_{\gamma} \begin{bmatrix} A_{11} & A_{12} \\ A_{21} & A_{22} \end{bmatrix} \begin{bmatrix} \alpha_{\gamma m}(0) \\ \beta_{\gamma m}(0) \end{bmatrix}, \quad (3)$$

where $F_{\gamma}$ is expressed as

$$F_e = \frac{\mu_0 ce_m(q_0;\eta)}{ce_m(q_{eff};\eta)} [J_{em}(q_{eff};\xi_0) H_{em}^{(1)\prime}(q_{eff};\xi_0) - J_{em}^{\prime}(q_{eff};\xi_0) H_{em}^{(1)}(q_{eff};\xi_0)]^{-1},$$

$$F_o = \frac{\mu_0 se_m(q_0;\eta)}{se_m(q_{eff};\eta)} [J_{om}(q_{eff};\xi_0) H_{om}^{(1)\prime}(q_{eff};\xi_0) - J_{om}^{\prime}(q_{eff};\xi_0) H_{om}^{(1)}(q_{eff};\xi_0)]^{-1}$$

and the entries of the matrix are

$$A_{11} = \mu_0 J_{\gamma m}(q_0;\xi_0) H_{\gamma m}^{(1)\prime}(q_{eff};\xi_0) - \mu_{eff} J_{\gamma m}^{\prime}(q_0;\xi_0) H_{\gamma m}^{(1)}(q_{eff};\xi_0), \quad (4a)$$

$$A_{12} = \mu_0 H_{\gamma m}^{(1)}(q_0;\xi_0) H_{\gamma m}^{(1)\prime}(q_{eff};\xi_0) - \mu_{eff} H_{\gamma m}^{(1)\prime}(q_0;\xi_0) H_{\gamma m}^{(1)}(q_{eff};\xi_0), \quad (4b)$$

$$A_{21} = \mu_{eff} J_{\gamma m}^{\prime}(q_0;\xi_0) J_{\gamma m}(q_{eff};\xi_0) - \mu_0 J_{\gamma m}(q_0;\xi_0) J_{\gamma m}^{\prime}(q_{eff};\xi_0), \quad (4c)$$

$$A_{22} = \mu_{eff} H_{\gamma m}^{(1)\prime}(q_0;\xi_0) J_{\gamma m}(q_{eff};\xi_0) - \mu_0 H_{\gamma m}^{(1)}(q_0;\xi_0) J_{\gamma m}^{\prime}(q_{eff};\xi_0). \quad (4d)$$

The effective medium condition requires that the scattering of the coated cylinder in the effective medium vanishes. Since the scattered field of the coated cylinder is represented by $H_m^{(1)}(q_{eff};\xi)$, a vanishing scattered wave implies $\beta_{\gamma m}(eff) = 0$. According to Eq. (3), such a condition leads to:

$$\frac{A_{21}}{A_{22}} = -\frac{\beta_{\gamma m}(0)}{\alpha_{\gamma m}(0)} = -D_{\gamma m}(0), \quad (4)$$

where $D_{\gamma m}(0)$ represent the Mie scattering coefficients of a scatterer of the metamaterial and can be obtained by solving the Helmholtz equation and matching the boundary conditions between the scatterer and the background medium. They are expressed as:

$$D_{\gamma m}(0) = \frac{\mu_0 J'_{\gamma m}(q_s;\xi_s)J_{\gamma m}(q_0;\xi_s) - \mu_s J_{\gamma m}(q_s;\xi_s)J'_{\gamma m}(q_0;\xi_s)}{\mu_s J_{\gamma m}(q_s;\xi_s)H^{(1)'}_{\gamma m}(q_0;\xi_s) - \mu_0 J'_{\gamma m}(q_s;\xi_s)H^{(1)}_{\gamma m}(q_0;\xi_s)}, \tag{5}$$

in which the using of subscript "$s$" means that the quantities take the corresponding values of the scatterer and $\xi_s$ indicates the boundary of the scatterer.

When the wavelength in the effective medium is much larger compared to the size of the coated cylinder, i.e., $\frac{1}{2}k_{eff}(a_0 + b_0) \ll 1$, the scattering of the coated cylinder is then dominated by the monopolar ($m=0$) and dipolar ($m=1$) terms. Under this condition, we can approximate the zero- and first-order Mathieu functions associated with the effective medium by

$$J_{e0}(q_{eff};\xi_0) = 1 \;,\quad J'_{e0}(q_{eff};\xi_0) = \frac{v_1^2 - v_2^2}{2} \;,\quad J_{e1}(q_{eff};\xi_0) = \frac{v_1 + v_2}{2} \;,\quad J'_{e1}(q_{eff};\xi_0) = -\frac{v_1 - v_2}{2} \;,$$

$J_{o1}(q_{eff};\xi_0) = -\frac{v_1 - v_2}{2}$ and $J'_{o1}(q_{eff};\xi_0) = \frac{v_1 + v_2}{2}$, with $v_1 = \sqrt{q_{eff}}e^{-\xi_0}$ and $v_2 = \sqrt{q_{eff}}e^{\xi_0}$. Then by using these approximated Mathieu functions and plugging Eqs. (4c) and (4d) into Eq. (5), we obtain,

$$\frac{\varepsilon_{eff} + 2\varepsilon_0 \dfrac{J'_{e0}(q_0;\xi_0)}{k_0^2 a_0 b_0 J_{e0}(q_0;\xi_0)}}{\varepsilon_{eff} + 2\varepsilon_0 \dfrac{Y'_{e0}(q_0;\xi_0)}{k_0^2 a_0 b_0 Y_{e0}(q_0;\xi_0)}} = \frac{Y_{e0}(q_0;\xi_0)}{iJ_{e0}(q_0;\xi_0)}\frac{D_{e0}(0)}{1 + D_{e0}(0)}, \tag{7a}$$

$$\frac{\mu_{eff,y} - \mu_0 \dfrac{b_0 J_{e1}(q_0;\xi_0)}{a_0 J'_{e1}(q_0;\xi_0)}}{\mu_{eff,y} - \mu_0 \dfrac{b_0 Y_{e1}(q_0;\xi_0)}{a_0 Y'_{e1}(q_0;\xi_0)}} = \frac{Y'_{e1}(q_0;\xi_0)}{iJ'_{e1}(q_0;\xi_0)} \frac{D_{e1}(0)}{1+D_{e1}(0)}, \quad (7b)$$

and

$$\frac{\mu_{eff,x} - \mu_0 \dfrac{J_{o1}(q_0;\xi_0)a_0}{J'_{o1}(q_0;\xi_0)b_0}}{\mu_{eff,x} - \mu_0 \dfrac{Y_{o1}(q_0;\xi_0)a_0}{Y'_{o1}(q_0;\xi_0)b_0}} = \frac{Y'_{o1}(q_0;\xi_0)}{iJ'_{o1}(q_0;\xi_0)} \frac{D_{o1}(0)}{1+D_{o1}(0)}, \quad (7c)$$

where $Y_{\gamma m}(q_0;\xi_0)$ stand for the Mathieu Neumann functions. Similar to the results derived for isotropic media [18], the effective permittivity and permeability are determined by monopolar ($m=0$) mode and dipolar ($m=1$) modes, respectively. However for the anisotropic case discussed here, the effective permeability is no longer a scalar, but a diagonalized tensor. It is interesting to notice that the elements of the tensor exactly correspond to the scattering coefficients of the first-order even and odd modes.

In the quasi-static limit, i.e. $\frac{1}{2}k_{eff}(a_0+b_0) \ll 1$, $\frac{1}{2}k_0(a_0+b_0) \ll 1$ and $\frac{1}{2}k_s(a_s+b_s) \ll 1$, the Mathieu functions in Eq. (7) can be approximated in the same way as that used to obtain Eq. (7). The Mathieu Neumann functions, $Y_{\gamma m}(q_\chi;\xi_\chi)$, and its derivatives, $Y'_{\gamma m}(q_\chi;\xi_\chi)$, can also be approximated as $Y_{e0}(q_\chi;\xi_\chi) = \frac{2}{\pi}\ln(v_{2\chi})$, $Y'_{e0}(q_\chi;\xi_\chi) = \frac{v_1^2}{\pi}\ln(v_{2\chi}) + \frac{2}{\pi}$,

$Y_{e1}(q_\chi;\xi_\chi) = \frac{v_{1\chi}}{\pi}\ln(v_{2\chi}) - \frac{2}{\pi v_{2\chi}}$, $Y'_{e1}(q_\chi;\xi_\chi) = \frac{2-v_{1\chi}^2}{\pi v_{2\chi}} + \frac{2v_{2\chi}-v_{1\chi}}{\pi}\ln(v_{2\chi}) + \frac{v_{1\chi}}{\pi}$,

$Y_{o1}(q_\chi;\xi_\chi) = -\frac{2}{\pi v_{2\chi}} - \frac{v_{1\chi}}{\pi}\ln(v_{2\chi})$ and $Y'_{o1}(q_\chi;\xi_\chi) = \frac{2-v_{1\chi}^2}{\pi v_{2\chi}} + \frac{2v_{2\chi}+v_{1\chi}}{\pi}\ln(v_{2\chi}) - \frac{v_{1\chi}}{\pi}$, with

$v_{1\chi} = \sqrt{q_\chi} e^{-\xi_\chi}$ and $v_{2\chi} = \sqrt{q_\chi} e^{\xi_\chi}$, where $\chi = 0,$ or $s$. Notice that $\frac{1}{2} k_0 (a_s + b_s) \ll 1$ is also satisfied in the quasi-static limit. Therefore, we can treat the corresponding Mathieu Bessel and Neumann functions as well as their derivatives, $J_{\gamma m}(q_0;\xi_s)$, $J_{\gamma m}'(q_0;\xi_s)$, $Y_{\gamma m}(q_0;\xi_s)$ and $Y_{\gamma m}'(q_0;\xi_s)$, in a similar way as we did previously. Eq. (7) can be then reduced to

$$\frac{\varepsilon_e - \varepsilon_0}{\varepsilon_0} = f \frac{\varepsilon_s - \varepsilon_0}{\varepsilon_0}, \tag{8a}$$

$$\frac{\mu_{eff,y} - \mu_0}{\frac{a_0}{a_0 + b_0}\mu_{eff,y} + \frac{b_0}{a_0 + b_0}\mu_0} = f \frac{\mu_s - \mu_0}{\frac{a_s}{a_s + b_s}\mu_s + \frac{b_s}{a_s + b_s}\mu_0}, \tag{8b}$$

and

$$\frac{\mu_{eff,x} - \mu_0}{\frac{b_0}{a_0 + b_0}\mu_{eff,x} + \frac{a_0}{a_0 + b_0}\mu_0} = f \frac{\mu_s - \mu_0}{\frac{b_s}{a_s + b_s}\mu_s + \frac{a_s}{a_s + b_s}\mu_0}, \tag{8c}$$

where $f = \pi a_s b_s / ab$ is the filling ratio of the elliptical cylinder. It is worth mentioning that Eq. (8) is nothing but exactly the recovery of Maxwell-Garnett (M-G) EMT [41], in which the effective parameters are functions of the filling ratio and do not depend on the frequencies.

In order to verify the validity of our EMT described in Eq. (7), we plot in black dots the band structures of a metamaterial obtained from a full-wave simulation in Fig. 2(a). The metamaterial is composed of rectangular array of elliptical cylinders embedded in air. The geometric sizes of the scatterer and the lattice are chosen as $a_s = 0.26r$, $b_s = 0.2r$, $a = 1.16r$, $b = 1.12r$, where $r$ is a normalized length unit, while the material parameters are chosen as $\varepsilon_s = 12, \mu_s = 1$ for the scatterer, and $\varepsilon_0 = 1$ $\mu_0 = 1$ for air. Also plotted in Fig. 2(a) in red solid curves are the band structures predicted by the EMT, i.e., Eq. (7). The corresponding effective permittivity and permeability are shown in Figs. 2(b) and 2(c), respectively, which offer us a clear picture in understanding of the dispersion relations. We mark three points on the band structures at the Brillouin zone center as points "A", "B", and "C" (the blue dots in Fig. 2(a)). The eigenfrequencies of these points are, respectively $\tilde{\omega}_A = 0.531$, $\tilde{\omega}_B = 0.555$, and $\tilde{\omega}_C = 0.593$, where dimensionless frequency $\tilde{\omega} = \omega a / 2\pi c_0$ is used ($c_0$ is the wave velocity in air.). Comparing Fig. 2(a) with Figs. 2(b) and 2(c), we find that $\tilde{\omega}_A$, $\tilde{\omega}_B$ and $\tilde{\omega}_C$ exactly correspond to the frequencies at which $\mu_{eff,y}$, $\varepsilon_{eff}$ and $\mu_{eff,x}$ turn to zero. Because the dispersion relations of such an anisotropic medium are determined by [7]:

$$\frac{k_{eff,x}^2}{\mu_{eff,y}} + \frac{k_{eff,y}^2}{\mu_{eff,x}} = \omega^2 \varepsilon_{eff}, \tag{9}$$

it is easy to deduce the dispersion relations along different directions. For example, in the ΓX (ΓY) direction, i.e., $k_{eff,y} = 0$ ($k_{eff,x} = 0$), we have $k_{eff,x} = \omega\sqrt{\varepsilon_{eff}\mu_{eff,y}}$ ($k_{eff,y} = \omega\sqrt{\varepsilon_{eff}\mu_{eff,x}}$). If both $\varepsilon_{eff}$ and $\mu_{eff,y}$ are positive (negative) over a frequency range, then there is a positive (negative) band in the ΓX direction. If one of them changes to different sign, then, a band gap in the ΓX direction rather than a pass band appears. The same rules apply to the dispersion relations along the ΓY direction if the role of $\mu_{eff,y}$ is replaced with $\mu_{eff,x}$. With these rules, all the dispersion relations near points "A", "B" and "C" can be easily interpreted. For example, for frequencies between $\tilde{\omega}_A$ and $\tilde{\omega}_B$, both $\varepsilon_{eff}$ and $\mu_{eff,x}$ are negative when $\mu_{eff,y}$ is positive. Therefore, there is a negative band along the ΓY direction, but a band gap along the ΓX direction. When the frequency is slightly higher and locates between $\tilde{\omega}_B$ and $\tilde{\omega}_C$, Figs. 2(b) and 2(c) give positive $\varepsilon_{eff}$ and $\mu_{eff,y}$, and negative $\mu_{eff,x}$, which explain the positive band along the ΓX direction and the band gap along the ΓY direction. The flat bands near Points "A" and "C" along the ΓY and ΓX directions are in fact the longitudinal bands induced by $\mu_{eff,y}$ and $\mu_{eff,x}$ equal to zero [18], respectively.

Figure 2(a) exhibits excellent agreements between the numerical simulations and the derived EMT in the center of the Brillouin zone. We also notice that the red curves deviate from the black dots when the Bloch wave vector is far away from the Γ point. This is reasonable because we used the approximation condition $\frac{1}{2}k_{eff}(a_0+b_0) \ll 1$ to obtain Eq. (7), which limits the range of applicability of the EMT. When the Bloch wave vector is so large that the condition no longer holds, the EMT deems to be inaccurate. Nevertheless, the derived EMT still gives good predictions of the effective medium parameters near the Γ point. Here, we emphasize that, for the current studied case, Eq. (7) is valid even when the dimensionless frequency is as high as 0.66, at which the wavelength in the background is $1.52a$ (or $1.57b$), *far beyond the quasi-static limit*. A systematic study of the applicability of the EMT will be presented in the Discussion section. Figures 2(d)-2(f) illustrate the electric field distributions of the eigenstates at points "A", "B" and "C", which clearly show an even dipolar mode, a monopolar mode, and an odd dipolar mode, respectively. These figures again support the results given by Eq. (7) that $\mu_{eff,y}$, $\varepsilon_{eff}$ and $\mu_{eff,x}$ are correspondingly determined by the scattering coefficients of even $m=1$ mode, $m=0$ mode and odd $m=1$ mode.

From Fig. 2, we can see that when the frequency takes values of $\tilde{\omega}_A$, $\tilde{\omega}_B$, and $\tilde{\omega}_C$, the system exhibits an anisotropic zero-index-material behaviors because one of the effective material parameters is near zero. Zero-index materials have bestowed with unprecedented abilities to manipulate electromagnetic waves [13, 15, 42-46]. Here, we would like to pay particular attention to frequency point $\tilde{\omega}_C$ where $\mu_{eff,x} = 0.0002 \to 0^+$, $\mu_{eff,y} = 0.5637 \gg \mu_{eff,x}$, and $\varepsilon_{eff} = 0.1175$. These effective medium parameters indicate that the system is an anisotropic zero-index material with only one component of the permeability tensor near zero. Very recently, such a medium was theoretically proposed and found to be capable of cloaking arbitrarily shaped defects and of exciting evanescent waves near the boundaries of the defects, which offers a new way to control the electromagnetic flux [15]. In the following, we provide simulated results of the wave transmission through such a metamaterial loaded with defects. The metamaterial is proposed based on our EMT predictions and can provide the same ability of cloaking. Figure 3(a) illustrates a schematic picture of the sample. It is an air waveguide filled with metamaterial slab (composed of $12 \times 10$ unit cells mentioned previously), inside which there distributed three defects "1", "2" and "3", as shown in Fig. 3(a), with sizes $2a \times 2b$, $2a \times b$ and $3a \times 2b$, respectively, and permeability $\mu = 1.5$, 0.4 and 2.1, respectively. The permittivity of the defects is simply set to 1. A TE-polarized plane wave with frequency $\tilde{\omega}_C$ is illuminated from the left.

As a comparison, we also plot in Fig. 3(b) the electric field for the same sample as shown in Fig. 3(a), but without the metamaterial slab. Notice that strong scattered waves are excited by the defects, which seriously distort the incident wave fronts. However, the results will be significantly altered in the presence of the metamaterial slab. Figures 3(c), 3(e) and 3(g) show, respectively, the electric field and the magnetic fields in the $x$ and $y$ directions. The field patterns at the outlet of the waveguide are found to be almost the same as the incident wave, showing the good cloaking effect of the metamaterial slab. Specifically, from Fig. 3(c), we clearly observe an almost uniform field distribution in the $y$-direction and an apparent phase change behavior along the $x$-direction in the metamaterial slab, implying that the metamaterial is highly anisotropic: the wavelength is nearly infinite long in the $y$-direction, but finite along the $x$-direction. The corresponding field distribution patterns for the same case but with the metamaterial slab replaced by its effective medium sample are plotted in Figs. 3(d), 3(f) and 3(h). Similar patterns to those shown in Figs. 3(c), 3(e) and 3(g) at the inlet and outlet of the waveguide are seen, suggesting that the EMT indeed describes the physical properties of the metamaterial. From Fig. 3(f), we find that evanescent waves around defects are induced within the horizontal regimes, which are essential for high transmittance [15].

Figure 3 demonstrates a functionality of the anisotropic zero-index metamaterial. Noting that the building blocks of the proposed metamaterial are dielectric elliptical cylinders, which are easily attainable, and that there are no complex structures and extremal material parameters involved, we believe that the fabrication of such a metamaterial is feasible.

Above, we support the validity and application of our anisotropic EMT with a simulated example, in which a set of values of $a_s/b_s$, $\varepsilon_s$, $\mu_s$ and filling ratio $f = \pi a_s b_s / ab$ are chosen. Good agreements between the numerical simulations and the EMT predictions are observed. In the following, we conduct a systematic study of how the material and geometric parameters influence the accuracy of the derived EMT. In Fig. 4, we plot the frequencies, at which zero effective medium parameters are obtained, as functions of various parameters. The curves are obtained from Eq. (7) and the dots correspond to the frequencies of the lowest monopolar and dipolar states at Γ point, which are results from the full-wave band-structure calculations. In Figs. 4(a)-4(c), under fixed permeability of the scatterers as 1, we change the aspect ratio, permittivity, and the filling ratio of the scatterers, respectively. In the lower panels of Fig. 4, we study the similar cases as those in the upper panels but with the permittivity of the scatterers fixed at 1. Figure 4 shows that the predictions of our EMT in general coincide with the band-structure simulations in a large parameters changing range. When the aspect ratio and filling ratio significantly increase, the predictions deviate from the numerical results. This is reasonable as higher angular momentum terms, i.e., $m \geq 2$, will contribute to the eigenmodes at low frequencies when the elliptical cylinder becomes flatter or bigger. This effect leads to inaccurate predictions because our effective medium scheme does not consider higher angular momentum terms.

In summary, we derive an anisotropic EMT for a 2D electromagnetic metamaterial, which provides *closed-form analytical* solutions for anisotropic effective medium parameters. The theory reveals the link between the effective medium parameters and the dominated resonant modes: the effective permittivity is related to the monopolar mode and the effective permeability tensor is associated with the even and odd dipolar modes. The validity of the derived EMT is verified by band-structure calculations. We find that the theory is valid even when the wavelength in the background medium is comparable to the size of the lattice, whose corresponding frequency locates beyond the long-wavelength limit. At the quasi-static limit, our EMT can recover the well-known Maxwell-Garnett formula. Based on these advantages, we expect our EMT will facilitate the design of new metamaterials. We also show by an example that a recently proposed anisotropic zero-index material can indeed be fabricated from a periodic structure. More anisotropic metamaterials with desired properties may also be devised from the predictions of our EMT. Although our theory is derived for electromagnetic metamaterials, it can be easily generalized to its acoustic counterparts because of the mathematical mapping between these two systems in two dimensions.

**Acknowledgements**

The work described here is supported by King Abdullah University of Science and Technology. The authors would like to thank Prof. P. Sheng, Prof. Z. Q. Zhang, Prof. J. Mei and Dr. M. Yang for stimulating discussions.

[24] M. G. Silveirinha, and P. A. Belov, *Phys. Rev. B* **77,** 233104 (2008).

[25] A. Alù, *Phys. Rev. B* **84**, 075153 (2011).

[26] A. Alù, *Phys. Rev. B* **83**, 081102(R) (2011).

[27] M. Yang, G. C. Ma, Y. Wu, Z. Y. Yang, and P. Sheng, *Phys. Rev. B* **89,** 064309 (2014).

[28] D. Torrent, and J. Sánchez-Dehesa, *New J. Phys.* **10,** 023004 (2008).

[29] O. Kidwai, S. V. Zhukovsky, and J. E. Sipe, *Phys. Rev. A* **85,** 053842 (2012).

[30] J. Mei, Y. Wu, and Z. Y. Liu, *Europhys. Lett.* **98,** 54001 (2012).

[31] D. Torrent, and J. Sánchez-Dehesa, *New J. Phys.* **13,** 093018 (2011).

[32] Y. Wu, J. Mei, and P. Sheng, *Physica B: Cond. Matt.* **407,** 4093 (2012).

[33] D. R. Smith, and J. B. Pendry, *JOSA B* **23,** 391 (2006).

[34] R. L. Chern, and Y. T. Chen, *Phys. Rev. B* **80,** 075118 (2009).

[35] Y. Lai, Y. Wu, P. Sheng, and Z. Q. Zhang, *Nat. Mater.* **10,** 620 (2011).

[36] D. R. Smith, S. Schultz, P. Markoš, and C. M. Soukoulis, *Phys. Rev. B* **65,** 195104 (2002).

[37] V. Fokin, M. Ambati, C. Sun, and X. Zhang, *Phys. Rev. B* **76**, 144302 (2007).

[38] X. X. Liu, D. A. Powell, and A. Alù, *Phys. Rev. B* **84**, 235106 (2011).

[39] X. X. Liu, and A. Alù, *Phys. Rev. B* **87**, 235136 (2013).

[40] J. C. Gutiérrez-Vega, R. M. Rodríguez-Dagnino, M. A. Meneses-Nava, and S. Chávez-Cerda, *Am. J. Phys.* **71,** 233 (2003).
18

**Figures**

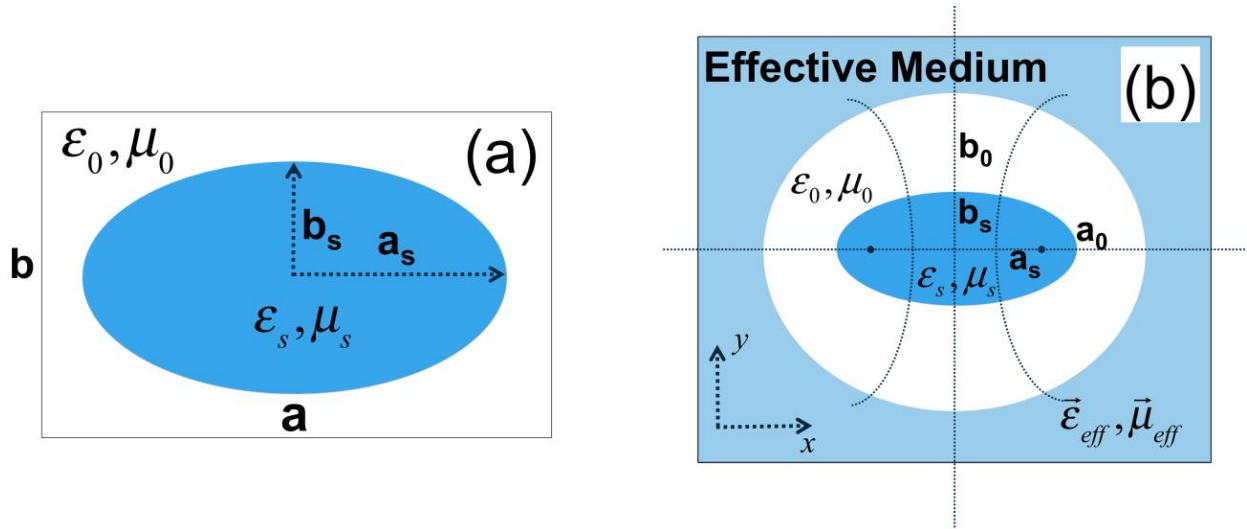

**Figure 1** (a) A schematic unit cell of proposed 2D periodic system which is composed of elliptical cylinders embedded in rectangular lattice. (b) The microstructure of the system.

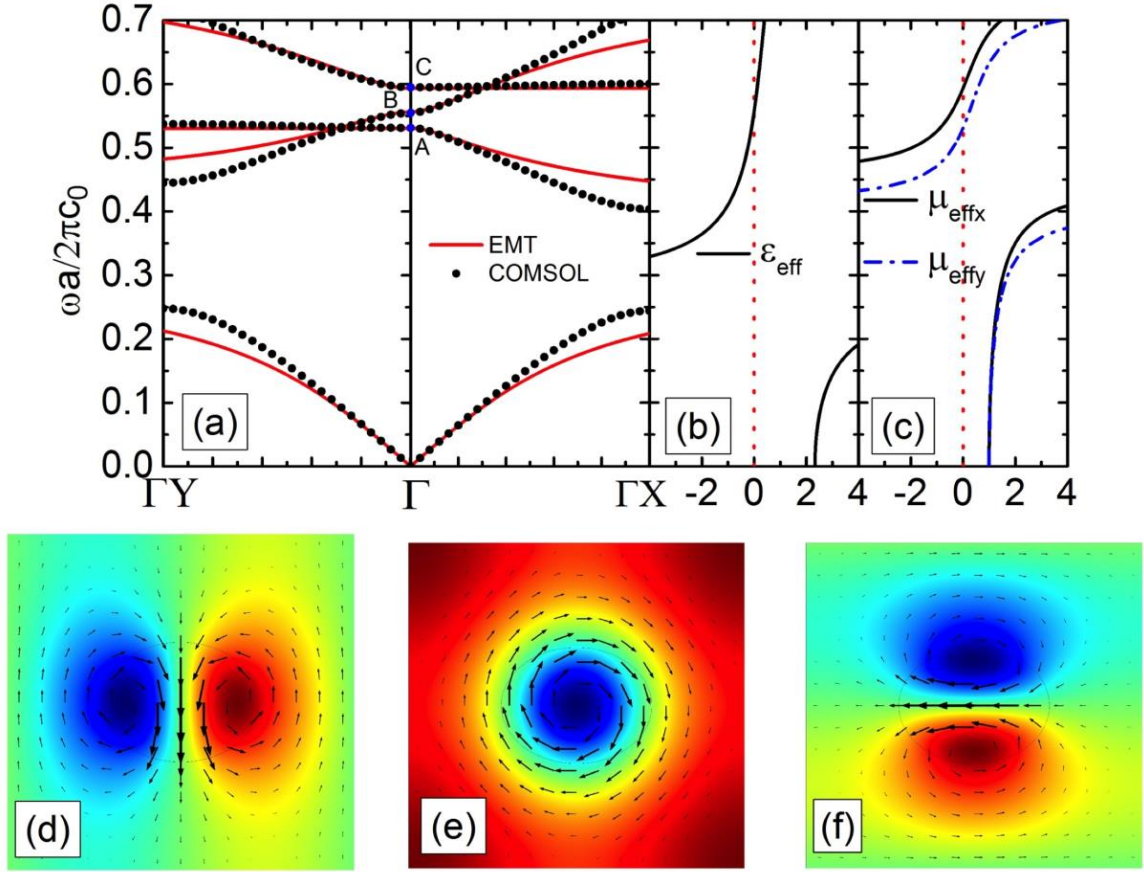

**Figure 2** Verification of the derived EMT. (a) Band structure calculations (black dots) using COMSOL, compared with EMT predictions (red curves) using Eq. (7). The geometric parameters are taken as $a_s = 0.26r$, $b_s = 0.2r$, $a = 1.16r$ and $b = 1.12r$. $\varepsilon_s = 12, \mu_s = 1$, $\varepsilon_0 = 1$ and $\mu_0 = 1$ are the material parameters. (b) The corresponding effective permittivity and (c) permeability calculated from Eq. (7). (d) Eigen electric field patterns for points "A", (e) "B" and (f) "C" marked in (a).

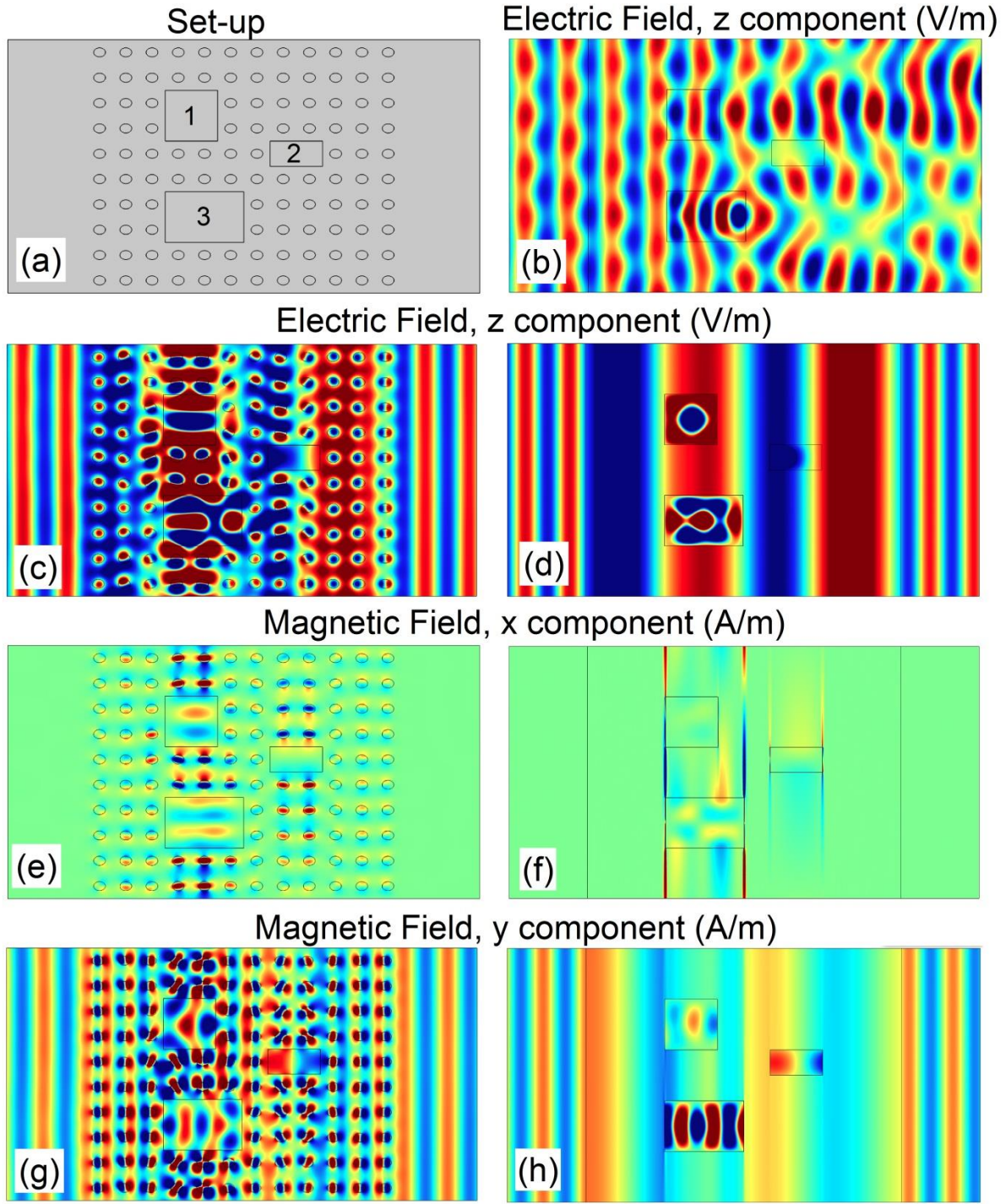

**Figure 3** Demonstration of wave transmission through an anisotropic zero-index metamaterial loaded with defects. (a) Schematic of the sample, which is an air waveguide filled with metamaterial slab (composed of 12×10 unit cells). Inside the metamaterial, there are three

defects marked as "1", "2" and "3", with sizes of $2a \times 2b$, $2a \times b$ and $3a \times 2b$, respectively, permeabilities $\mu = 1.5$, 0.4 and 2.1, respectively, and permittivity 1. (b) The electric field pattern for a TE-polarized plane wave with frequency $\tilde{\omega}_C = 0.593$ impinges from the left side of the waveguide without the metamaterial. (c) The electric field pattern, (e) the $x$-component of the magnetic field pattern, and (g) the $y$-component of magnetic field pattern under the same excitation condition as (b) but with the metamaterial slab in the waveguide. [(d) (f) and (h)] The same quantities as those described in (c), (e) and (g), respectively, but the metamaterial slab is replaced with an effective homogenous slab, which possesses effective medium parameters $\varepsilon_{eff} = 0.1175$, $\mu_{eff,x} = 0.0002$ and $\mu_{eff,y} = 0.5637$.

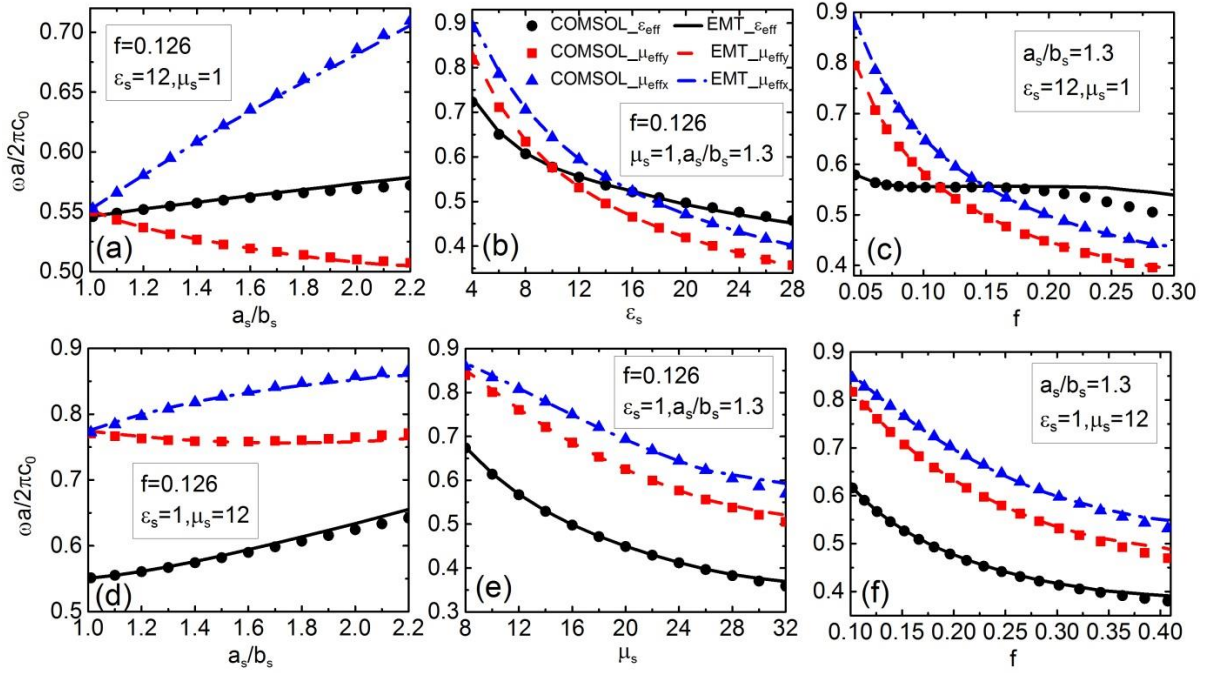

**Figure 4** The influences of different parameters on the predictions of the derived EMT. The frequencies at which $\varepsilon_{eff}$, $\mu_{eff,x}$ or $\mu_{eff,y}$ becomes zero predicted by Eq. (7) as functions of various parameters are pictured as curves. For comparison, the frequencies of the lowest monopolar and dipolar states at the Γ point, which are obtained from the band structure calculations using COMSOL, are also plotted in dots. (a) Influence of changing $a_s/b_s$ with fixed $\varepsilon_s = 12$, $\mu_s = 1$ and $f = 0.126$. (b) Influence of changing $\varepsilon_s$, with fixed $a_s/b_s = 1.3$, $\mu_s = 1$ and $f = 0.126$. (c) Influence of changing $f$, with fixed $a_s/b_s = 1.3$, $\varepsilon_s = 12$ and $\mu_s = 1$. [(d)-(f)] Similar studies as those shown in (a)-(c) but the dielectric cylinders ($\mu_s = 1$) are replaced with magneto cylinders ($\varepsilon_s = 1$).